\documentclass[11pt]{article}
\usepackage{color}
\usepackage{amsmath}
\usepackage{amssymb}
\usepackage{latexsym}
\usepackage{psfrag,fancyhdr,epsfig}

\begin{document}

\begin{titlepage}

\begin{centering}

\hfill hep-th/yymmnnn\\

\vspace{1 in}
{\bf {INFLATION WITH THE RIGHT-HANDED SNEUTRINO REVISITED} }\\
\vspace{1 cm}
{O. Efthimiou$^{1}$ and K. Tamvakis$^{1,2}$}\\
\vskip 0.5cm
{$^1 $\it{Physics Department, University of Ioannina\\
45110 Ioannina, GREECE}}\\
\vskip 0.5cm
{$^2$\it {Physics Department, CERN, CH-1211, Geneva 23,\\
Switzerland}}

\vspace{1.5cm}
{\bf Abstract}\\
\end{centering}
\vspace{.1in}
We consider an extension of $\nu MSSM$ with an extra $U(1)$ that realizes $D$-term inflation driven by the right-handed sneutrino. Non-renormalizable terms in the Kahler potential and the superpotential are considered, the latter controlled by a suitable discrete $R$-symmetry. We find that, for subplanckian inflaton values, the predictions of inflationary parameters are compatible with observations, establishing the right-handed sneutrino driven inflation as a viable scenario.

\vfill

\vspace{2cm}
\begin{flushleft}

December 2009
\end{flushleft}
\hrule width 6.7cm \vskip.1mm{\small \small}
 \end{titlepage}

The most dramatic development of the past decade in particle
physics, namely the discovery of neutrino mass \cite{N}, requires
the presence of a right-handed neutrino field, which, in the
supersymmetric version of the Standard Model, is accompanied by
its scalar partner, a {\textit{right-handed sneutrino}}. A
straightforward way to explain the smallness of the neutrino mass
is to invoke the {\textit{seesaw mechanism}} \cite{SS}, in which
the right-handed neutrino possesses a large mass in the range of
$10^{11}$ to $10^{15}\,GeV$. What is remarkable is that
right-handed sneutrino fields can be related to inflation, thus,
providing a direct connection between cosmological inflation and
particle physics. Models where the right-handed sneutrino is the
inflaton have been proposed, either in the chaotic inflation
framework \cite{SNUCI1,SNUCI2} or in the framework of $D$-term
inflation \cite{SNUSG}. Although the minimal sneutrino
inflationary scenarios, based on chaotic inflation, result in
density perturbations compatible with existing data, from the
point of view of particle physics, the model should be embedable
to supergravity. Then, one has to address the flatness of the
scalar potential, which, in general, is problematic due to
supergravity corrections of $F$-terms ({\textit{$\eta$-problem}}).
This can be circumvented if one adopts the $D$-term inflation
\cite{D1,D2} framework, where the required vacuum energy that
drives inflation is supplied by Fayet-Iliopoulos $D$-term of a
$U(1)$ gauge factor. An alternative approach is to employ a
specific form of the Kahler potential in order to maintain the
flatness of the scalar potential ($F$-term hybrid inflation) but
extra fields are required \cite{FHI,FHTFI}.

In the present article we reconsider inflation driven by the
right-handed sneutrino in the $D$-term inflation framework. We
consider an extension of the standard $\nu MSSM$ that realizes
neutrino masses through the see saw mechanism, with an extra
$U(1)$ gauge factor, under which all standard fields are neutral
apart from a pair of MSSM singlets $\phi_{\pm}$. Suitable
symmetries, such as $R$-parity and discrete $R$-symmetries,
restrict the superpotential couplings of these fields. We consider
leading non-renormalizable corrections to the superpotential and
the Kahler potential, while restricting ourselves to subplanckian
values of the inflaton field. We find that sneutrino driven
inflation leads to inflationary parameters and, in particular, the
spectral index, compatible with observations. Thus, modulo general
inflationary issues, such as the gravitino problem, right-handed
sneutrino driven inflation seems to be a viable scenario.

The extension of the MSSM with three right-handed neutrino superfields $N_i^c$ realizes the seesaw mechanism through the renormalizable superpotential
\begin{equation}
{\cal{W}}_0\,=\,\frac{M_R^{(i)}}{2}N_i^cN_i^c\,+\,Y_{ij}N_i^cL_jH^c\,.{\label{RENO}}
\end{equation}
This is the most general renormalizable superpotential, assuming $R$-parity conservation and taking the right-handed neutrino to be odd under it. We shall extend this model further by introducing an extra $U(1)$ gauge factor under which all fields are neutral except an oppositely charged pair $\phi_+$ and $\phi_-$. We are also going to assume that a non-zero Fayet-Iliopoulos $D$-term is present. A renormalizable superpotential term $N^c\phi_+\phi_-$ would lead to a neutrino state of electroweak mass but, here, it is forbidden by $R$-parity, if we take the product of $\phi_{\pm}$ to be even. Then, an extra symmetry would be necessary in order to forbid the direct $\phi_+\phi_-$ mass-term. A suitable such symmetry is {\textit{a discrete $R$-symmetry}} and as a specific example we may take the following ${\cal{Z}}_3^{(R)}$ symmetry\footnote{All standard terms of the superpotential are allowed, provided $(H^c,\,H,\,L,\,E^c,\,Q,\,D^c,\,U^c)\,\rightarrow\,(1,\,\alpha^2,\,\alpha,\,\alpha^2,\,\alpha,\,\alpha^2,\,\alpha)$.
This symmetry is not assumed to be a symmetry of the sector of the theory responsible for the spontaneous breaking of local supersymmetry. It will be explicitly broken by supersymmetry breaking.}
\begin{equation}N_i^c\,\rightarrow\,\alpha\,N_i^c,\,\,\,\phi_{\pm}\,\rightarrow\,\phi_{\pm},\,\,\,\,{\cal{W}}\,\rightarrow\,\alpha^2 {\cal{W}}\,.\end{equation}
Adopting this symmetry, we see that at the lowest non-renormalizable level the only allowed term is $\frac{\lambda_{ij}}{M}N_i^cN_j^c\phi_+\phi_-\,$. Also, there are no allowed terms of $O(M^{-2})$. Actually, the allowed superpotential form, involving only right-handed neutrino fields and $\phi_{\pm}$,  to all orders can be written down as
\begin{equation}{\cal{W}}_{N}\,=\,\left(N^c\right)^2\,{\cal{F}}\left(\phi_+\phi_-,\,\left(N^c\right)^6\,\right)\,=\,\sum_{n,\,m\,=\,0}\frac{C_{n,\,m}}{M^{2m+6n-1}}\left(\phi_+\phi_-\right)^m\left(N^c\right)^{3n+2}\,.\end{equation}
The first few allowed terms, apart from ({\ref{RENO}}), are
$$
\,\frac{\lambda_{ij}}{M}N_i^cN_j^c\phi_+\phi_-\,+\,\frac{\lambda_{ij}'}{M^3}N_i^cN_j^c\left(\phi_+\phi_-\right)^2\,+\,\frac{\lambda_{ij}''}{M^5}N_i^cN_j^c\left(\phi_+\phi_-\right)^3
+\,\frac{\lambda_{i_1\dots i_8}}{M^5}N_{i_1}^c\,\dots\,N_{i_8}^c\,+\,\dots\,\,.$$
Throughout this article we assume that the defining scale of non-renormalizable terms will be of the order of the reduced Planck mass and we set $M\,=\,M_P\,\sim\,2.4\times 10^{18}\,GeV$. The only other dimensionfull parameters appearing in the superpotential are the right-handed neutrino masses $M_R^{(i)}$. In order to obtain an acceptable neutrino mass through the seesaw mechanism, we must take the right-handed neutrino masses $M_1,\,M_2,\,M_3$ in the range $10^{10}-10^{14}\,GeV$. We shall assume that one of these right-handed sneutrino fields, namely, the lightest, will play a role in inflation and we shall suppress family indices in what follows.

Since, we have considered non-renormalizable terms in the
superpotential, we must do the same with the Kahler potential as
well \cite{SJ}. Restricting ourselves to the quadratic
$\phi_{\pm}$ term, we have
\begin{equation}
{\cal{K}}\,=\,|N^c|^2\,f(n)\,+\,g(n)\,\left(|\phi_+|^2\,+\,|\phi_-|^2\right)\,+\,\dots\,,{\label{KAL}}
\end{equation}
where $f(n)$ and $g(n)$ are arbitrary functions defined as
\begin{equation}\,f(n)\,=\,\sum_{j=0}f_jn^j,\,\,\,\,\,\,\,\,\,g(n)\,=\,\sum_{j=0}g_jn^j\,\,\,\,\,\,\,\,\,\,n\,\equiv\,\frac{|N^c|^2}{M^2}\,.\end{equation}
 The dots in ({\ref{KAL}}) correspond to higher powers of $\phi_{\pm}$, which will be neglected, since we anticipate that $\phi_{\pm}$ will either stay at the origin or obtain values $<<M$.

We may next proceed to calculate the scalar potential resulting from ({\ref{KAL}}) and the superpotential
\begin{equation}
{\cal{W}}\,=\,\frac{M_R}{2}{N^c}^2\,\,+\,\frac{\lambda}{2M}{N^c}^2\left(\phi_+\phi_-\right)\,+\,\dots\,,{\label{SUPER}}
\end{equation}
where the dots denote terms of $O(M^{-3})$ or higher. The $F$-term part of the potential is
 $$V_F\,\approx\,e^{{\cal{K}}/M^2}\left\{\,{\cal{F}}_i\left({\cal{K}}_j^{j}\right)^{-1}{\cal{F}}^j\,-3\frac{|{\cal{W}}|^2}{M^2}\,\right\}\,\approx\,$$
$$\frac{|N^c|^2}{A}\left[M_R^2+\frac{\lambda^2}{M^2}|\phi_+\phi_-|^2+\frac{\lambda M_R}{M}\left(\phi_+\phi_-+\phi_+^*\phi_-^*\right)+\frac{M_R^2}{2M^2}\left({N^c}^2+{{N^c}^*}^2\right)(f+nf^{(1)})\,\right]\,$$
$$-\frac{BM_R^2}{M^2A^2}|N^c|^2\left(|\phi_+|^2+|\phi_-|^2\right)\,+\,\,\,\frac{\lambda^2|N^c|^4}{4g(n)M^2}\left(|\phi_+|^2+|\phi_-|^2\right)\,+\,$$
\begin{equation}\frac{M_R^2|N^c|^2}{AM^2}\left[\,f(n)|N^c|^2+g(n)\left(|\phi_+|^2+|\phi_-|^2\right)\,\right]\,
-\frac{3M_R^2}{4M^2}|N^c|^4\,+\,\dots\,.\end{equation}
We have assumed that field values will be below $M$ and kept terms up to $O(M^{-2})$, while keeping the functions $f$ and $g$ intact. The functions $A$ and $B$ stand for
\begin{equation}A(n)\,\equiv\,f(n)\,+\,3nf^{(1)}(n)\,+\,n^2f^{(2)}(n),\,\,\,\,\,\,\,\,B(n)\,\equiv\,g^{(1)}\,+\,n\,g^{(2)}(n)\,.\end{equation}
If we ignore the subleading terms proportional to $M_R$, we obtain
\begin{equation}
V_F  \approx {{\lambda ^2 \left| {N^c } \right|^2 } \over {AM^2
}}\left| {\phi _ +  \phi _ -  } \right|^2  + {{\lambda ^2 \left|
{N^c } \right|^4 } \over {4M^2 g(n)}}\left( {\left| {\phi _ +  }
\right|^2  + \left| {\phi _ -  } \right|^2 } \right)\,+\,\dots
\,.\end{equation}
The $D$-term part of the scalar potential is
\begin{equation}V_D\,=\,\frac{\tilde{g}^2}{2}\left({\cal{K}}_{+}\phi_+\,-{\cal{K}}_-\phi_-\,+\,\xi\right)^2\,=\,\frac{\tilde{g}^2}{2}\left(g(n)\left(|\phi_+|^2-|\phi_-|^2\right)\,+\,\xi\right)^2\,,\end{equation}
where we have introduced a Fayet-Iliopoulos $D$-term $\xi\,>>\,M_R^2$.

The $\phi_{\pm}$-mass terms read off from the potential are
\begin{equation}M_{\pm}^2\,=\,\frac{\lambda^2}{4M^2g(n)}\,|N^c|^4\,\pm\tilde{g}^2\xi\,g(n)\,.{\label{KRIT}}\end{equation}
Both masses are positive for
\begin{equation}n^2\,=\,\frac{|N^c|^4}{M^4}\,\geq\,\frac{4\tilde{g}^2}{\lambda^2}\,\frac{\xi}{M^2}\,g^2(n)\,.{\label{nnn}}\end{equation}
Anticipating subplanckian values for $N^c$, we may define the
{\textit{critical field}}
\begin{equation}
n_c\,\equiv\,\frac{2\tilde{g}}{\lambda}\,g_0\,\frac{\sqrt{\xi}}{M}\,,
\end{equation}
where $g(n)\,\approx\,g_0\,+\,n\,g_1\,+\,\dots.$. Thus,
({\ref{nnn}}) becomes just $n\,\geq\,n_c$, or, in terms of a real
field
\begin{equation}
N^c\,=\,\frac{\phi}{\sqrt{2}}\,\Longrightarrow\,\phi\,\geq\,\,\phi_c\,=\,2\sqrt{\frac{g_0\tilde{g}}{\lambda}\sqrt{\xi}\,M}\,.\end{equation}
For $\phi\,>\,\phi_c$ both $\phi_{\pm}$ have positive masses and their vevs stay at the origin.  In the unbroken phase the scalar potential is
$$V(\phi)\,\approx\,\frac{\tilde{g}^2\xi^2}{2}\,+\,O(M_R^2)\,\phi^2\,+\,O(M_R^2/M^2)\,\phi^4\,+\,\dots\,,$$
which is approximaterly equal to $\tilde{g}^2\xi^2/2$ and very flat. Thus, when the Universe is in the above global vacuum the energy density is constant and inflation can occur. The amount of inflation depends on the initial and final value the sneutrino field which plays the role of the inflaton. When the critical value $\phi\,\sim\,\phi_c$ is reached the above local minimum ceases to exist and the Universe makes a transition to the global minimum
\begin{equation}
\phi_+\,=\,0,\,\,\,\,\phi_-\,\approx\,\sqrt{\xi/g_0}\,.
\end{equation}
and inflation stops.

At the local vacuum $\phi_{\pm}=0$, the tree potential receives considerable radiative corrections given by the standard Coleman-Weinberg formula in the terms of the split masses of the $\phi_{\pm}$ superfields as\footnote{The kinetic terms, near the origin $\phi_{\pm}=0$, are
$$\,A(n)\,|\partial_{\mu}N^c|^2\,+\,g(n)(|D_{\mu}\phi_+|^2+|D_{\mu}\phi_-|^2)\,.$$}
$$\Delta V\,=\,\frac{1}{32\pi^2}\sum_{\pm}\,\frac{M_{\pm}^4}{g^2(n)}\ln\left(\frac{M_{\pm}^2}{g(n)\Lambda^2}\right)\,
-\frac{1}{16\pi^2}\frac{M^4(0)}{g^2(n )}\ln\left(\frac{M^2(0)}{g(n)\Lambda^2}\right)\,,$$
where
$$M_{\pm}^2\,=\,M^2(0)\,\pm\,\tilde{g}^2\xi\,g(n)\,\,.$$
Thus, we finally obtain
\begin{equation}V\,\approx\,\frac{\tilde{g}^2}{2}\xi^2\,+\,\frac{\tilde{g}^4\xi^2}{16\pi^2}\,\ln\left(\frac{\phi^4}{g^2(n)\tilde{\Lambda}^4}\right)\,,{\label{RAD}}\end{equation}
where we have suitably chosen the cutoff to absorb the constant factors.

Let us conclude the presentation of the model by discussing the
mass-scales involved. We have already assumed that the scale of
the right-handed neutrino mass $M_R$ is much smaller than the
other scales involved. The assumed range of $M_R$ is between
$10^{10}$ and $10^{14}\,GeV$. Apart from the reduced Planck scale
$M\,\sim\,2.4\times\,10^{18}\,GeV$, the only other scale appearing
is the Fayet-Iliopoulos scale $\sqrt{\xi}$. There is a well known
constraint for $\xi$ coming from the formation of cosmic strings
\cite{KSI}, namely
\begin{equation}
3.8\times 10^{15}\,GeV\,\leq\,\sqrt{\xi}\,\leq\,4.6\times 10^{15}\,GeV\,.{\label{STRING}}
\end{equation}
In order to safely ignore contributions from the right-handed neutrino mass $M_R$, we should have
$$M_R^2\,\phi^2\,<<\,\tilde{g}^2\xi^2\,\,\Longrightarrow\,M_R\,<<\,\tilde{g}\,\xi/\phi_{max}\,.$$
Taking $\sqrt{\xi}\,\sim\,4.6\times 10^{15}GeV$ and $\tilde{g}\,\sim\,O(0.1)$, for $\phi_{max}\,\sim\,M$, we obtain
$$M_R\,<<\,10^{12}\,GeV\,.$$
Note that the non-minimality of the Kahler potential will not modify this significantly, since, for subplanckian values of $n$, $A(n)\,\approx\,f_0$ and the physical mass is just $M_R^*\,=\,M_R/\sqrt{f_0}$ with $f_0\,\sim\,O(1)$.

 In the local vacuum the scalar potential is well approximated by the constant $\tilde{g}^2\xi^2/2$ plus the radiative corrections part ({\ref{RAD}}). Assuming that the slow-roll approximation is valid, namely $\ddot{\phi}<<H\,\dot{\phi}$ and $(\dot{\phi})^2<<V(\phi)$, the classical evolution equations are
 \begin{equation}
 3H\,\dot{\phi}\,\approx\,-\frac{V'(\phi)}{A(\phi)},\,\,\,\,\,\,H^2\,\approx\,\frac{V(\phi)}{3M^2}\,\Longrightarrow\,\frac{d\phi}{d\ln a}\,\approx\,-\frac{M^2}{A(\phi)}\frac{V'(\phi)}{V(\phi)}\,.
 \end{equation}
 The above expression can be integrated to give the {\textit{number of e-folds}} ${\cal{N}}$ as
 \begin{equation}
 {\cal{N}}\,=\,\ln\left(\frac{a_f}{a_i}\right)\,\approx\,\frac{1}{M^2}\int_{\phi_f}^{\phi_i}d\phi\,A(\phi)\,\frac{V(\phi)}{V'(\phi)}\,=\,\frac{2\pi^2}{\tilde{g}^2}\int_{n_f}^{n_i}dn\left(\frac{f(n)+3nf^{(1)}(n)+n^2f^{(2)})n)}{1-n\frac{g^{(1)}}{g}}\right)\,.{\label{INT}}
 \end{equation}
 Anticipating $n<<1$, we approximate and obtain
 $$
 {\cal{N}}\,\approx\,\frac{2\pi^2}{\tilde{g}^2}\int_{n_f}^{n_i}dn\left(\,f_0\,+\,n\left(4f_1+\frac{g_1}{g_0}f_1\right)\,+\,\dots\right)\,$$
 or
 \begin{equation}\frac{\tilde{g}^2}{2\pi^2}{\cal{N}}\approx\,\left(\,f_0(n_i-n_f)\,+\,\frac{1}{2}\left(n_i^2-n_f^2\right)\left(4f_1+\frac{g_1}{g_0}f_0\right)\,\right)\,.{\label{EF}}
 \end{equation}

\begin{figure}[t]
  \begin{center}
  \includegraphics[width = 0.6 \textwidth] {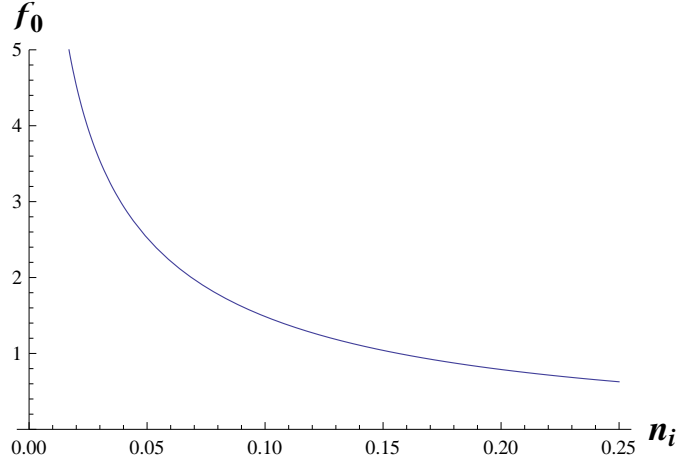}
    \caption{ The Kahler coefficient $f_0$ as a function of $n_i$. }
   \label{f0}
  \end{center}
\end{figure}

The {\textit{comoving curvature perturbation}}, in terms of the
potential ({\ref{RAD}}), is
$$
{\cal{R}}_c\,=\,\frac{H^2}{2\pi|\dot{\phi}|}\,\approx\,\frac{V^{3/2}A}{2\pi\sqrt{3}M^3|V'|}\,=\,\frac{\pi}{\sqrt{2}\sqrt{3}\tilde{g}}\left(\frac{\xi}{M^2}\right)\left(\frac{\phi_i}{M}\right)\left(\frac{f(n_i)+3n_if^{(1)}(n_i)+n_i^2f^{(2)}(n_i)}{1-n_i\frac{g^{(1)}(n_i)}{g(n_i)}}\right)\,$$
\begin{equation}\,\approx\,\frac{\pi}{\sqrt{6}\tilde{g}}\left(\frac{\xi}{M^2}\right)\left(\frac{\phi_i}{M}\right)\left(f_0\,+\,n_i\left(4f_1+\frac{g_1}{g_0}f_0\right)\,\right)\,.\end{equation}
Matching this to the observed value
${\cal{R}}_c\,\approx\,4.7\times 10^{-5}\,$ and choosing
$\sqrt{\xi}\,\approx\,4.6\times 10^{15}\,GeV$ and
$\tilde{g}\,\approx\,0.1$, amounts to the constraint
 \begin{equation}\sqrt{n_i}\left(f_0\,+\,n_i\left(4f_1+\frac{g_1}{g_0}f_0\right)\,\right)\,\approx\,0.705\,.
 \end{equation}
 Similarly, ({\ref{EF}}), for ${\cal{N}}\,\sim \,65$, can be thought off as an equation constraining $n_f$.

 As a very rough approximation, we shall assume that $f_1\,\sim\,f_0$ and $g_1\,\sim\,g_0$. The range of values for $f_0$ is shown in fig.1 for a range of subplanckian values of $n_i$. Assuming that inflation ends when the value $n_c$ is reached, we may identify
\begin{equation}n_f\,\approx\,n_c\,=\,\frac{2\tilde{g}}{\lambda}\,g_0\,\frac{\sqrt{\xi}}{M}\,.\end{equation}
For the chosen values, of $\sqrt{\xi}\,\sim\,4.6\times
10^{15}\,GeV$ and $\tilde{g}\,\sim\,0.1$, this corresponds to
$n_c\,\sim\,0.38\times 10^{-3}\,(g_0/\lambda)$ and requires a
small (but not unnatural ) coupling $\lambda\,\sim\,g_0\,O(0.01)$.
In fig.2 $n_f$ is plotted as a function of $n_i$.

\begin{figure}[t]
  \begin{center}
  \includegraphics[width = 0.6 \textwidth] {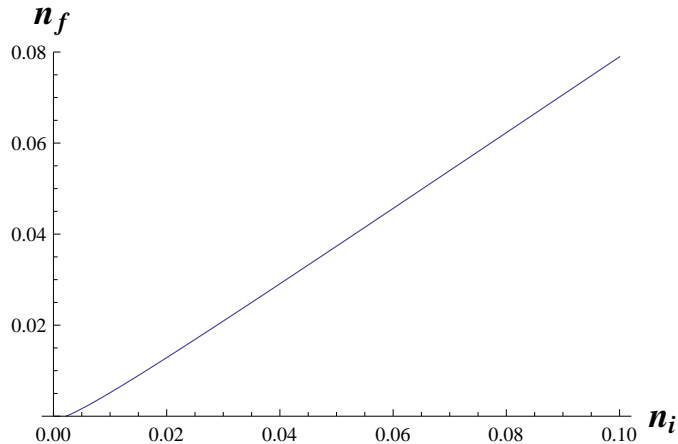}
    \caption{ $n_f$ as a function of $n_i$. }
   \label{nf}
  \end{center}
\end{figure}

Let us now consider the {\textit{slow-roll parameters}}
$\epsilon,\,\eta$ and $\zeta$. They are given in terms of the
potential and its derivatives. We have
\begin{equation}\epsilon\,=\,M^2\left(\frac{V'}{V}\right)^2\,\approx\,\frac{1}{8n_i}\left(\frac{\tilde{g}}{\pi}\right)^2\left(1-2n_i\frac{g_1}{g_0}\,+\,O(n_i^2)\right)
\end{equation}
\begin{equation}\eta\,=\,2M^2\frac{V''}{V}\,\approx\,-\frac{1}{2n_i}\left(\frac{\tilde{g}}{\pi}\right)^2\left(1+n_i\frac{g_1}{g_0}\,+\,O(n_i^2)\,\right)
\end{equation}
\begin{equation}\zeta\,=\,\frac{M^2}{2}\sqrt{\frac{V'''V'}{V^2}}\,\approx\,\frac{1}{4\sqrt{2}n_i}\left(\frac{\tilde{g}}{\pi}\right)^2\left(1\,-n\frac{g_1}{g_0}\,+\,O(n_i^2)\,\right)\,.
\end{equation}

The corresponding {\textit{spectral index}} $n_s$ is
\begin{equation}
n_s\,=1+2\eta_i-6\epsilon_i\,\approx\,1+2\eta_i\,\approx\,1\,-\frac{1}{n_i}\left(\frac{\tilde{g}}{\pi}\right)^2\left(1+n_i\frac{g_1}{g_0}\,+\,O(n_i^2)\,\right)
\end{equation}
and it is plotted in fig.3 for the choice $\tilde{g}\,\sim\,0.1$
and $g_0\,\sim\,g_1$. From this plot one can immediately see that
our values for the spectral index are compatible with the
corresponding value from observational data,
$n_s\,=\,0.963^{+0.014}_{-0.015}$ \cite{Dunkley}.
\begin{figure}[t]
  \begin{center}
  \includegraphics[width = 0.7 \textwidth] {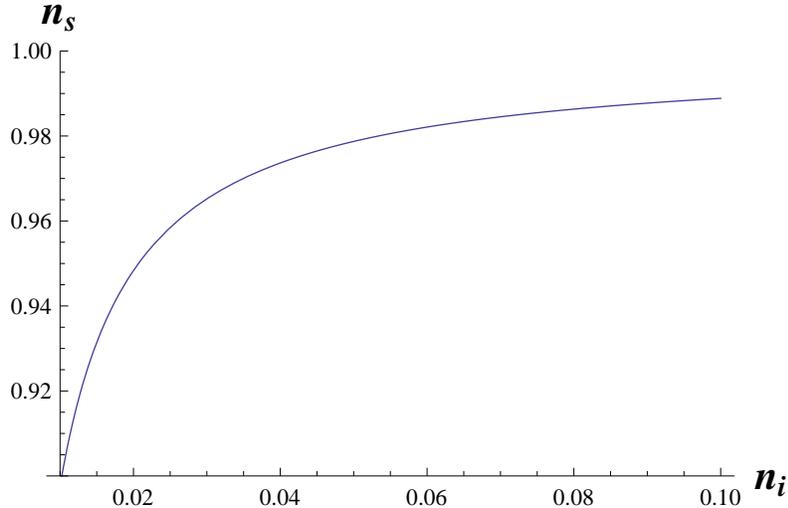}
    \caption{ Spectral index $n_s$ as a function of $n_i$. }
   \label{f0}
  \end{center}
\end{figure}

We have assumed that the right-handed sneutrino that drives inflation is the lightest one. It is well known that this sneutrino through its lepton violating processes can produce the desired lepton number asymmetry. Its decays ($\tilde{N}^c\rightarrow\,\tilde{L}^*\,+\,{H^c}^*$)  will reheat the Universe to a temperature $T_R\,=\,1.4\times 10^{10}\,\sqrt{M_R^*(GeV)/10^{10}}\sqrt{\sum_j|Y_{1j}|^2/10^{-6}}\,.$
In order to avoid overproduction of gravitinos this temperature must not exceed $10^6-10^7\,GeV$. This corresponds to the preferred values $Y\sim 10^{-6}$ and $M_N\,\sim\,10^{10}\,GeV$. Note however that it is possible that a late time entropy production will dilute the gravitino density without the need of small Yukawa couplings. Apart from the ubiquitous gravitino problem, there is also another issue, namely the problem associated with the decays of the heavy fields $\phi_{\pm}$, a typical problem of all models of $D$-term inflation. These decays can lead to a potentially high reheating temperature causing overproduction of gravitinos. Note however that the subsequent decay of the right-handed sneutrino at a lower temperature will produce additional entropy that can sufficiently dilute gravitinos. The precise way this can occur depends on the range of various parameters.

Let us now conclude summarizing the main points of this article. We have considered the right-handed neutrino extension of the MSSM that realizes neutrino masses through the see saw mechanism. We extended this model further with a $U(1)$ gauge group under which all standard fields are neutral apart from a pair of MSSM singlets $\phi_{\pm}$. Suitable symmetries, such as $R$-parity and discrete $R$-symmetries, restrict the superpotential couplings of these fields to a class of non-renormalizable operators. As it stands the model can realize the scenario of $D$-term inflation with the inflaton being identified with the right-handed sneutrino field. Considering also non-minimal corrections to the Kahler potential but staying in the subplanckian field space, we arrive at inflationary predictions compatible with existing data, thus, establishing the possibility of right-handed sneutrino driven inflation as a viable scenario.

{\textbf{Acknowledgements}}

The authors acknowledge the support of European Research and
Training Networks MRTPN-CT-2006 035863-1 (UniverseNet) and "UNILHC" PITN-GA-2009-237920 (Unification in the LHC era). K.T. thanks the CERN Theory Group for its hospitality.

\end{document}